\documentclass[reprint,prb,superscriptaddress,showpacs]{revtex4-2}%
\usepackage{graphicx}
\usepackage{color}
\usepackage{amsmath}
\usepackage{amsfonts}
\usepackage{amssymb}
\usepackage{framed}
\usepackage{float}
\usepackage[normalem]{ulem}
\providecommand{\U}[1]{\protect\rule{.1in}{.1in}}
\makeatletter
\renewcommand*{\fnum@figure}{{\normalfont\bfseries \figurename~\thefigure}}
\renewcommand*{\@caption@fignum@sep}{\textbf{ : }}
\makeatother
\usepackage{hyperref}
\hypersetup{
    colorlinks=true,
    linkcolor=blue,
    filecolor=magenta,      
    urlcolor=cyan,
}
\begin{document}

\title{Magnetism and Fermiology of Kagome Magnet YMn$_6$Sn$_4$Ge$_2$}

\author{Hari Bhandari}
\email{hbhandar@gmu.edu}
\affiliation{Department of Physics and Astronomy, George Mason University, Fairfax, VA 22030, USA}
\affiliation{Quantum Science and Engineering Center, George Mason University, Fairfax, VA 22030, USA}
\author{Rebecca L. Dally}
\affiliation{NIST Center for Neutron Research, National Institute of Standards and Technology, Gaithersburg, MD 20899, USA}
\author{Peter E. Siegfried}
\affiliation{Department of Physics and Astronomy, George Mason University, Fairfax, VA 22030, USA}
\affiliation{Quantum Science and Engineering Center, George Mason University, Fairfax, VA 22030, USA}
\author{Resham B. Regmi}
\affiliation{Department of Physics and Astronomy, George Mason University, Fairfax, VA 22030, USA}
\affiliation{Quantum Science and Engineering Center, George Mason University, Fairfax, VA 22030, USA}
\author{Kirrily C. Rule}
\affiliation{Institute for Superconducting and Electronic Materials, The University of Wollongong, Australia}
\affiliation{Australian Nuclear Science and Technology Organisation, New Illawarra Road, Lucas Heights, NSW, 2234, Australia}
\author{Songxue Chi}
\affiliation{Neutron Scattering Division, Oak Ridge National Laboratory, Oak Ridge, TN 37831, USA}
\author{Jeffrey W. Lynn}
\affiliation{NIST Center for Neutron Research, National Institute of Standards and Technology, Gaithersburg, MD 20899, USA}
\author{I. I. Mazin}
\affiliation{Department of Physics and Astronomy, George Mason University, Fairfax, VA 22030, USA}
\affiliation{Quantum Science and Engineering Center, George Mason University, Fairfax, VA 22030, USA}
\author{Nirmal J. Ghimire}
\email{nghimire@gmu.edu}
\affiliation{Department of Physics and Astronomy, George Mason University, Fairfax, VA 22030, USA}
\affiliation{Quantum Science and Engineering Center, George Mason University, Fairfax, VA 22030, USA}

\date{\today}
\begin{abstract}
Kagome lattice magnets are an interesting class of materials as they can host topological properties in their magnetic and electronic structures. YMn$_6$Sn$_6$ is one such compound in which a series of competing magnetic phases is stabilized by an applied magnetic field, and both an enigmatic topological Hall effect and a Dirac crossing close to the Fermi energy have been realized. This material also shows a magnetization-induced Lifshitz transition and evidence of a unique charge-spin coupling in one of the magnetic phases, namely the fan-like phase. Tuning the magnetism, and thus the interplay with the electronic states, opens new avenues for precise control of these novel properties. Here, we demonstrate the extreme sensitivity of the magnetic phases in YMn$_6$Sn$_4$Ge$_2$ through the investigation of structural, magnetic, and transport properties. The high sensitivity to small doping  provides great potential for engineering the magnetic phases and associated electronic properties in this family of rare-earth kagome magnets.
\end{abstract}
\maketitle

\section{Introduction}\label{sec:1}

Materials exhibiting exotic magnetic textures in the presence of nontrivial electronic structures have shaped the direction of much research in recent years. Investigations into these materials have led to many predictions and experimental observations of quantum phenomena including the quantum anomalous Hall effect \cite{liu2016quantum,chang2013experimental}, chiral Majorana modes \cite{akhmerov2009electrically, daido2017majorana}, and the topological magnetoelectric effect \cite{lee2015manifestation, wu2016quantized}. These phenomena rely on the nature of the electronic structure in the presence of time-reversal symmetry breaking driven by the material's magnetism \cite{Haldane1988,Ohgushi2000,Armitage2018}. An increasingly popular class of materials to investigate the interplay of topological magnetic and electronic structures are the kagome-net magnets \cite{zhang2011quantum, ghimire2020topology,kang2020dirac,Ye2018c,yin2019negative,kuroda2017evidence}.

RT$_6$Sn$_6$ compounds, where R represents a rare-earth element, and T is a 3d transition metal element,  are the recent addition to the class of kagome magnets \cite{ghimire2020competing,Hasitha2022,Elliott2022, pokharel2021electronic,connor2022origin,Shuang2021PRL,Ganesh2022PRM}. In these compounds, the T atoms are arranged in the kagome geometry, and the crystal structure provides a rich materials space to tune both the electronic and magnetic properties. Compounds with non-magnetic R atoms are simpler as magnetism comes only from the T sublattice. The most widely studied, and arguably the most interesting member of this family, is YMn$_6$Sn$_6$ where multiple exotic properties have been realized including a series of competing magnetic phases and the topological Hall effect \cite{ghimire2020competing,dally2021chiral}, Dirac bands \cite{li2021dirac,li2022manipulation}, a rarely observed magnetization-driven Lifshitz transition \cite{siegfried2022magnetization}, a large anomalous transverse thermoelectric effect \cite{roychowdhury2022large}, and an emergent electromagnetic induction \cite{kitaori2021emergent}. 

\begin{figure}[ht!]
\begin{center}
\includegraphics[width=1\linewidth]{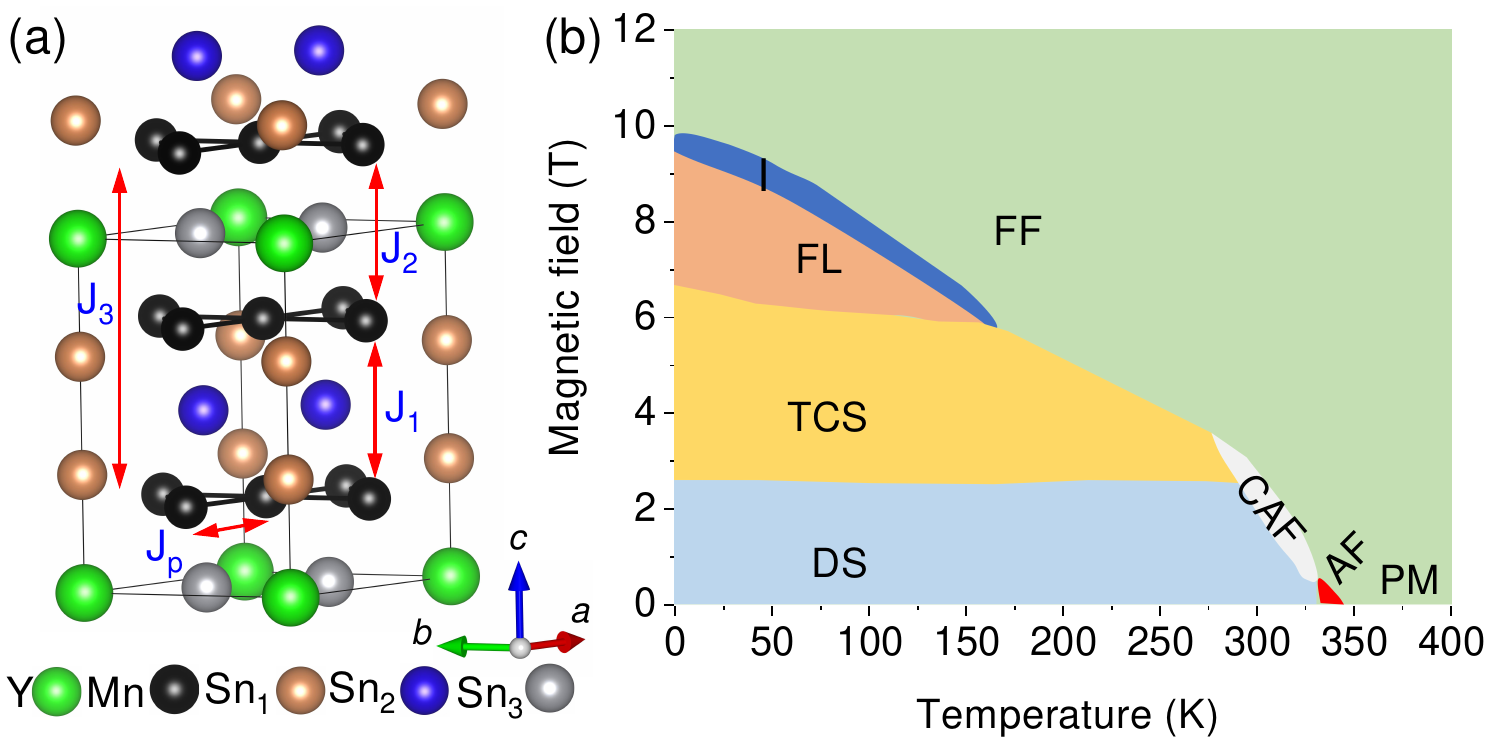}
    \caption{\small Crystal structure (a) Sketch of the crystal structure of YMn$_6$Sn$_6$  (b) and in-plane magnetic phase diagram of YMn$_6$Sn$_6$. J$_s$ in panel (a) represent the magnetic exchange interactions in different Mn planes indicated by the red lines.}
    \label{structure}
    \end{center}
\end{figure}

The magnetic structure drives most of these phenomena in YMn$_6$Sn$_6$ and the interesting magnetism in this compound comes from a parametric frustration facilitated by its crystal structure shown in Fig. \ref{structure}(a) \cite{ghimire2020competing}. It consists of Mn kagome layers in the $ab$-plane that are separated alternately by two distinct blocks along the $c$-axis: a Sn$_2$R layer and a Sn$_3$ layer. In the first approximation, the energy of this system can be described by the Hamiltonian \cite{ghimire2020competing}:

\begin{align*}
\mathcal{H}=&{\displaystyle\sum\limits_{i,j}}J_{n}\mathbf{n}_{i}\cdot
\mathbf{n}_{j}+{\displaystyle\sum\limits_{i,j}}J_{p}\mathbf{n}_{i}
\cdot\mathbf{n}_{j}\\
+&{\displaystyle\sum\limits_{i}}J^{z}n_{i}^{z}\cdot n_{i+1}^{z}%
+K{\displaystyle\sum\limits_{i}}(n_{i}^{z})^{2}
+{\displaystyle\sum\limits_{i}}\mathbf{n}_{i}\cdot\mathbf{H,}\label{H}%
\end{align*}
where $\mathbf{H}$ is the external field and $\mathbf{n}$ is a unit vector
along the local magnetization direction. The first sum runs over the three nearest neighbors along the $c$-axis, the second over the first nearest neighbors in the $ab$-plane, the next over all vertical bonds ($i$+1 denotes the nearest $c$-neighbor), and last two over each Mn ion. $K$ is the single-ion anisotropy, and the Ising-type anisotropic exchange, $J^{z}$, is the only one allowed by symmetry for the vertical bonds. As a further simplification, in our previous paper \cite{ghimire2020competing} we limited the latter term to nearly-ferromagnetic vertical bonds only.

The in-plane interaction among the Mn atoms within a kagome layer is strongly ferromagnetic ($J_p<0$). The spins are forced to lie in the $ab$-plane \cite{ghimire2020competing, VENTURINI1996102} due to the net effect of the single-ion term (easy-axis) and the Ising exchange (easy-plane and stronger). The most interesting part of the magnetic order is the helical magnetic spiral at $T=0$ and $H=0$ which comes from the competition between $J_1$, $J_2$, and $J_3$, and is therefore 
very sensitive to the stacking  pattern of the Mn-layers along the $c$-axis. In YMn$_6$Sn$_6$, the exchange interactions through two inequivalent layers along the $c$-axis within a unit cell are opposite in sign. Specifically [Fig. \ref{structure}(a)], the interaction across the Sn$_3$ layer is ferromagnetic (FM) ($J_1<0$) and that across the Sn$_2$Y is antiferromagnetic ($J_2>0$). The exchange 
interaction between the like-Mn layers (i.e., the next-nearest neighbor interaction) in YMn$_6$Sn$_6$ is weak and ferromagnetic $J_3<0$. In the absence of $J_3$, a commensurate antiferromagnetic structure is expected \cite{dally2021chiral}. In fact, this is the case just below the N\'eel temperature ($T_N$) of 345 K. The presence of $J_3$ introduces parametric frustration and the system transitions into a unique staggered spiral phase below 333 K that persists down to the lowest measured temperatures\cite{ghimire2020competing}.  

The application of an external magnetic field in the $ab$-plane stabilizes a series of competing magnetic phases. A sketch of the magnetic field (B)-temperature (T) phase diagram of YMn$_6$Sn$_6$ obtained in previous studies \cite{ghimire2020competing,dally2021chiral} is depicted in Fig. \ref{structure}(b). At low temperatures there are four main magnetic phases, namely distorted spiral (DS), transverse conical spiral (TCS), fan-like (FL), and forced ferromagnetic (FF). In addition, there are three smaller phases labelled ``I", the canted antiferromagnetic phase (CAF), and the antiferromagnetic phase (AF).  

As these various magnetic structures in YMn$_6$Sn$_6$ are directly related to the crystal structure, mainly set by the spacing blocks between the Mn layers, replacement of Sn atoms by isoelectronic Ge provides a clean way to influence the magnetic properties by tuning these exchange interactions. In this work, we report the synthesis of YMn$_6$Sn$_4$Ge$_2$ and the study of its magnetic and magnetotransport properties by means of magnetic and transport measurements, and neutron diffraction experiments; and, by means of first principles calculations, we provide the microscopic insight into the influence of the magnetic structure on its electronic structure and thus the transport properties. We find that in YMn$_6$Sn$_4$Ge$_2$, Ge replaces Sn preferentially from one of the three sites [see structural details in Section \ref{sec:2}(A) and Table \ref{T1}]. Directly related to this particular replacement, YMn$_6$Sn$_4$Ge$_2$ orders directly into the incommensurate spiral phase below $T_N$ of 345 K with the spins lying in the $ab$-plane, and gives rise to a distinct magnetic and electronic structure. Magnetization and magnetotransport measurements show marked differences from that of the parent compound and indicate that new magnetic phases are stabilized. Counterintuitively, conductivity is enhanced, more significantly along the $c$-axis due to the change in fermiology introduced by the Ge doping. Our results provide an important insight into the doping study of not only YMn$_6$Sn$_6$, but all RMn$_6$Sn$_6$ compounds that are currently attracting significant attention for electronic and magnetic topological states.

\section{Results}\label{sec:2}

\subsection{ Crystal chemistry}

\begin{figure}[ht!]
\begin{center}
\includegraphics[width=.8\linewidth]{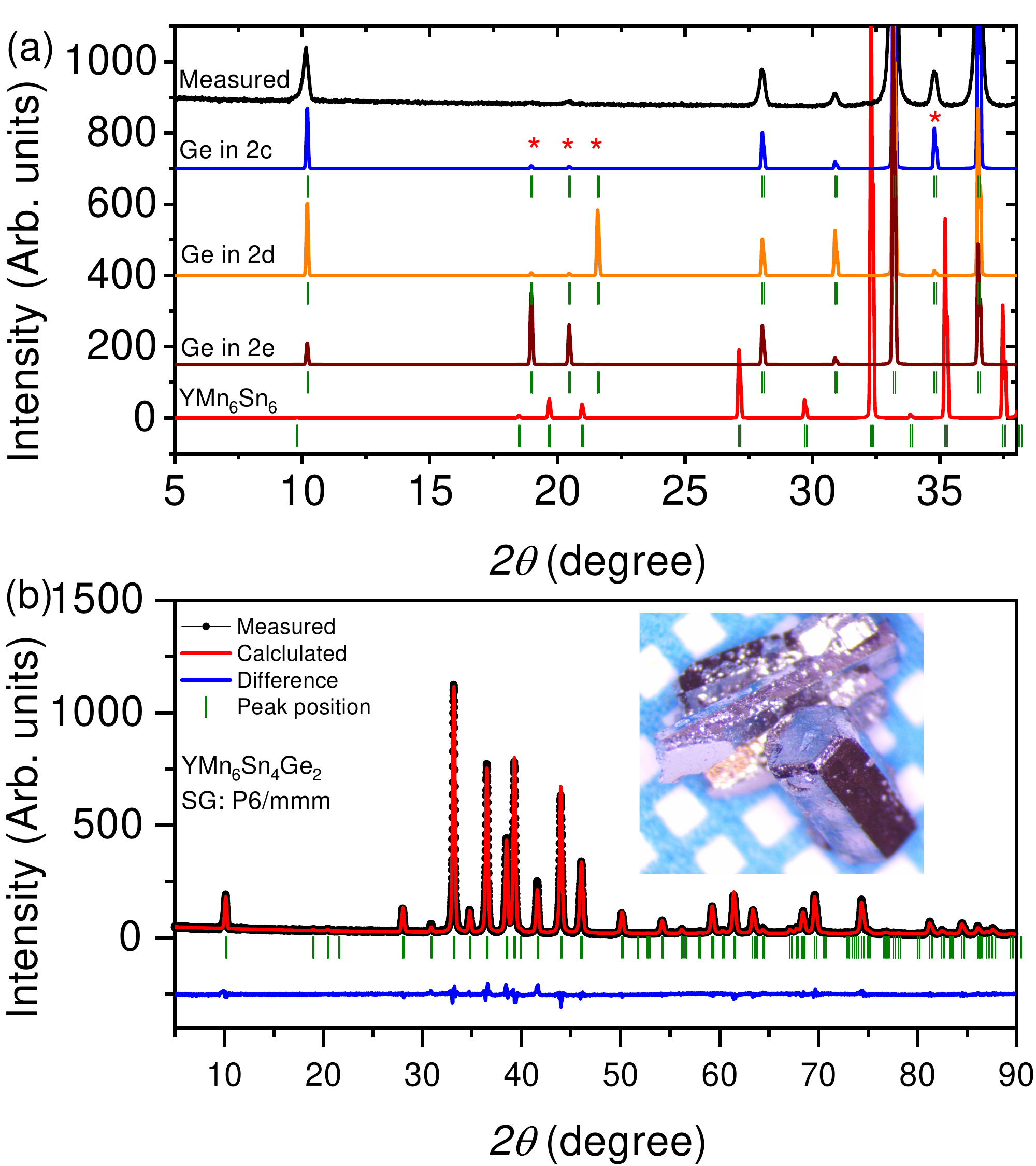}
    \caption{\small (a) Calculated powder x-ray diffraction patterns of YMn$_6$Sn$_6$ and YMn$_6$Sn$_4$Ge$_2$ with Ge in the three crystallographic sites 2c, 2d and 2e plotted together with the measured room temperature x-ray powder pattern of YMn$_6$Sn$_4$Ge$_2$. The same lattice parameters were used in the calculation for the three structures of YMn$_6$Sn$_4$Ge$_2$. Here, the x-ray powder patterns are shown between $5^{\circ} \leq 2\theta \leq 37^{\circ}$. Asterisks are used to emphasize some diffraction peaks. (b) Rietveld refinement of room temperature powder x-ray diffraction pattern of YMn$_6$Sn$_4$Ge$_2$. The model with Ge in the 2c position is used for the refinement. Inset shows an optical image of YMn$_6$Sn$_4$Ge$_2$ single crystals on a 2 mm $\times$ 2 mm grid (blue lines on the background). The cylindrical morphology of YMn$_6$Sn$_4$Ge$_2$ single crystals is markedly different than that of the parent compound which grow in a plate-like shape. The crystallographic $c$-axis in these crystals is along the length of the cylinder.}
    \label{Xray}
    \end{center}
\end{figure}

YMn$_6$Sn$_6$ crystallizes in the HfFe$_6$Sn$_6$-type structure in the hexagonal space group $P6/mmm$ (\#191). The crystallographic data taken from Ref. \cite{ghimire2020competing} are presented in Table \ref{T1}. In this structure, there are three inequivalent Sn positions where Sn$_1$, Sn$_2$, and Sn$_3$ take Wyckoff positions 2e, 2d, and 2c, respectively. Previous doping studies have reported the dopant atoms entering into different Sn positions. For example, in YMn$_6$Sn$_{6-x}$Ga$_x$, Ga enters into the 2c site \cite{lefevre2002neutron}, while in YMn$_6$Sn$_{6-x}$In$_x$, In takes the 2d site \cite{lefevre2003neutron}. An analysis of calculated x-ray diffraction patterns of YMn$_6$Sn$_4$Ge$_2$ with Ge in each of the three positions, as depicted in Fig. \ref{Xray}(a), shows that the intensity distribution in each of these cases is distinct, which makes it easier to identify the position taken by Ge in our sample. In Fig. \ref{Xray}(a) we plot the experimental powder x-ray diffraction pattern of YMn$_6$Sn$_4$Ge$_2$ together with the calculated patterns in different possible structures. The intensity distribution of the Bragg peaks marked by asterisks clearly shows that Ge in this compound takes the 2c position. This analysis is further verified by the Rietveld refinement of the powder x-ray pattern as shown in Fig. \ref{Xray}(b). The results of the refinement are presented in Table \ref{T2}. The best refinement result was obtained for 93\% of Ge and 7\% of Sn at the 2c site. For simplicity, we use the formula YMn$_6$Sn$_{4}$Ge$_2$ throughout this article and the same is used in the analysis of magnetic measurements as it does not make a noticeable difference. The refinement shows that the YMn$_6$Sn$_4$Ge$_2$ lattice parameters $a$ and $c$ are smaller than in the parent compound, with the $c$ ($a$) axis shrinking by 3.8\% (2.5\%). It is to be noted that it is the Sn in the Sn$_2$Y layer that is replaced by Ge in YMn$_6$Sn$_4$Ge$_2$. This essentially influences $J_2$ and $J_3$, but leaves $J_1$ unaffected [see Fig. \ref{structure}(a)]. 

This result is in agreement with our DFT calculations. In the latter, we first fully optimized the crystal structure with one Sn replaced by Ge in each of the three crystallographically inequivalent Sn sites. We found a very strong preference for the 2c position and the calculated total energy per formula was 0.35 eV (0.48 eV) lower than for the 2d (2e) position (see Table S1 in Supplementary Materials). Next, we have  fully optimized the structure of YMn$_6$Sn$_4$Ge$_2$ with Ge in the 2c position. The results are shown in Table \ref{T2} together with our experimentally obtained structure.

\begin{table}[h]
\caption{Crystallographic data of YMn$_6$Sn$_6$. The lattice parameters obtained at room temperature are: $a = b = 5.5398$ \AA\ and $c = 9.0203$ \AA\ \cite{ghimire2020competing}.}\label{T1}
\begin{center}
\par%
\begin{tabular}
[c]{c@{\hspace{0.3cm}}c@{\hspace{0.3cm}}c@{\hspace{0.3cm}}c@{\hspace{0.3cm}}c@{\hspace{0.3cm}}c}\hline
 Atom & Position & x & y & z                    \\
 \hline
Y   &   1a  &   0   &   0   &   0   \\
 Mn  &  6i   &   1/2   &   0   &   0.24587   \\
Sn$_1$    &  2e   &   0   &   0   &   0.33679   \\
Sn$_2$    &  2d   &   1/3   &   2/3   &   1/2   \\
Sn$_3$    &  2c   &   1/3   &   2/3   &   0   \\

 \hline
\end{tabular}
\end{center}
\end{table}

\begin{table}[h]
\caption{Selected data from Rietveld refinement of powder
x-ray diffraction collected on ground crystals of YMn$_6$Sn$_4$Ge$_2$, and from our DFT optimization.}\label{T2}
\centering
\begin{tabular}
{c|c|c}
 \hline
Space group & \multicolumn{2}{c}{\it P6/mmm} \\
 & exp.& calc.\\
\hline
 Unit cell $a$, $c$ (\AA) & 5.3993(7),~8.6797(13)& 5.339,~8.691\\
 \textit{R}$_{WP}$                     &     12.0 \%                               \\
  \textit{R}$_{B}$                     &     5.74 \%                               \\
   \textit{R}$_{F}$                           &     6.12  \%                               \\
\end{tabular}
\begin{tabular}
{cc|cccc|ccc}
		\hline	
                     &               & \multicolumn{4}{c|}{exp.}&\multicolumn{3}{c}{calc.}\\
        Atom         & Position      & $x$         &	  $y$	          &	      $z$	         &	  Occupancy	     & $x$         &	  $y$	          &	      $z$	      \\
		\hline
 Y & 1$a$ & 0 & 0 & 0 & 1.00& 0 & 0 & 0  \\														
 Mn & 6$i$ & 1/2 & 0 & 0.23245 & 1.00& 1/2 & 0 & 0.2330  \\ 
 
Sn$_1$ & 2$e$ & 0 & 0 & 0.33709 & 1.00& 0 & 0 & 0.3342  \\
 
Sn$_2$ & 2$d$ & 1/3 & 2/3 & 1/2 & 1.00  & 1/3 & 2/3 & 1/2 \\

Sn$_3$ & 2$c$ & 1/3 & 2/3 & 0 & 0.07  \\

Ge & 2$c$ & 1/3 & 2/3 & 0 & 0.93 & 1/3 & 2/3 & 0 \\

  \hline      					
\end{tabular}

\label{T-2}
\end{table}

\subsection{Magnetic properties}

 \begin{figure}[ht]
\begin{center}
\includegraphics[width=1\linewidth]{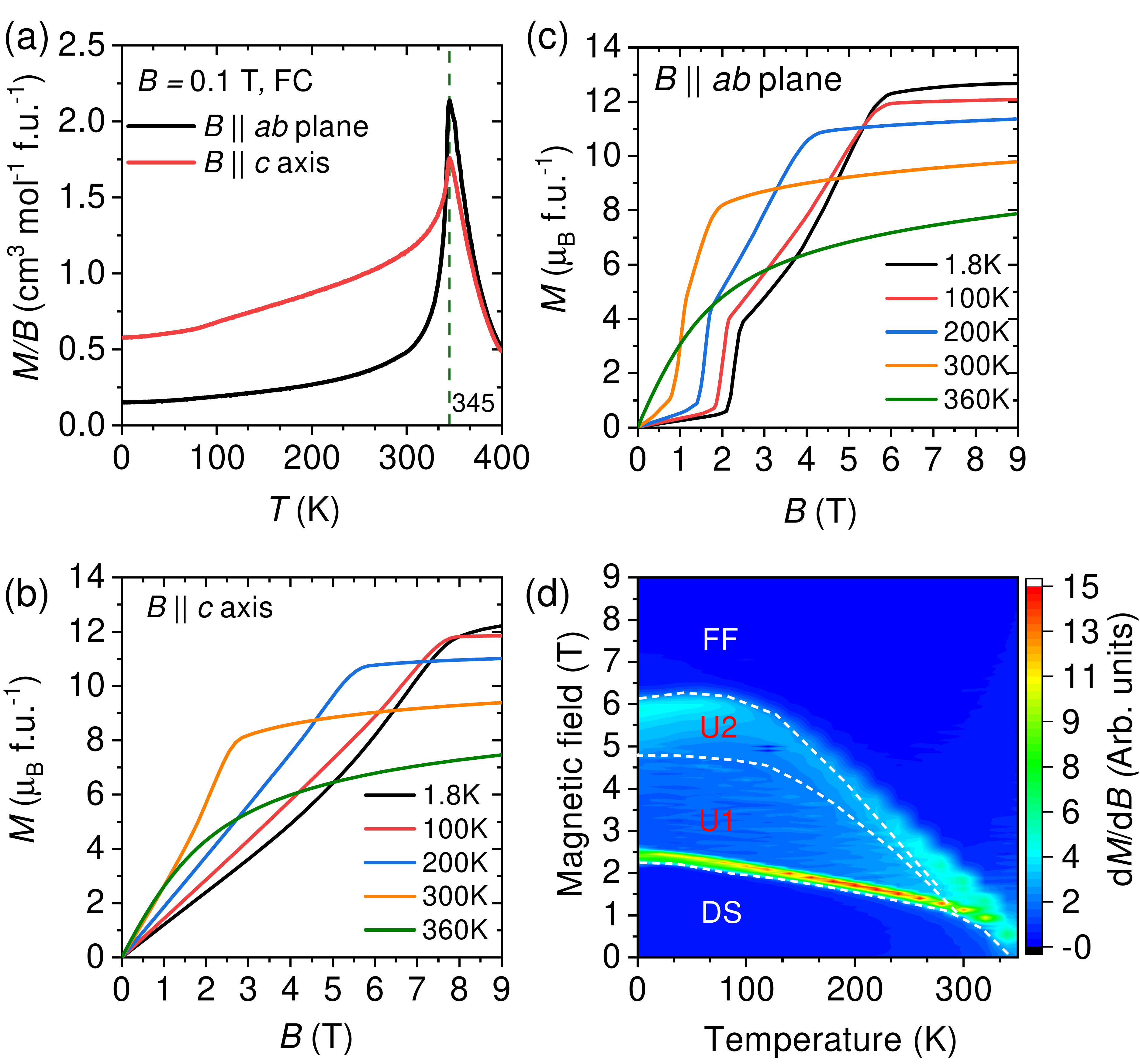}
\caption{\small (a) Magnetic susceptibility ($M/B$) of YMn$_6$Sn$_4$Ge$_2$ measured with a magnetic field $B = 0.1$ T, parallel and perpendicular to the $c$-axis. Field-cooled (FC) protocol was used for these measurements. (b,c) Magnetization $M$ of YMn$_6$Sn$_4$Ge$_2$ as a function of external magnetic field ($B$) applied along the $c$-axis, and in the $ab$-plane at selected temperatures. (d) Magnetic phase diagram of YMn$_6$Sn$_4$Ge$_2$ for a magnetic field applied in the  $ab$-plane, constructed from a color contour plot of d$M$/d$B$ data measured between 1.8 K and 360 K every 20 K. The white dashed lines are guides to the eye at phase boundaries. DS and FF are distorted spiral and forced ferromagnetic phases, respectively. U1 and U2 are two unknown magnetic phases.}
    \label{MT}
    \end{center}
\end{figure}
 DC magnetic susceptibility ($M/B$, where $M$ is the magnetic moment and $B$ is the external magnetic field) of YMn$_6$Sn$_4$Ge$_2$ for $B$ = 0.1 T, as a function of temperature ($T$) is depicted in Fig. \ref{MT}(a). The black (red) curve is the susceptibility measured with $B$ parallel to the $ab$-plane ($c$-axis). Both of these curves peak at 345 K indicating that the magnetic ordering takes place below this temperature. 
 
 Magnetization ($M$ vs $B$) data of YMn$_6$Sn$_4$Ge$_2$ for $B \| c$-axis at some representative temperatures are presented in Fig. \ref{MT} (b). At 1.8 K, $M$ increases gradually with increasing $B$, and saturates above 8 T, attaining the saturated moment of 12.2 $\mu_B$ per formula unit. This behavior of $M$ remains the same in the entire temperature range below $T_N$ with the exception that the saturation field and the saturated moment (as expected) decrease with increasing temperature. Above $T_N$ (at 360 K), the $M$ vs $B$ curve shows a rather monotonic increase, which indicates the presence of spin correlations above $T_N$, expected in this class of materials, as described in Ref. \cite{ghimire2020competing}. 
 
 Magnetization data measured with the magnetic field in the $ab$-plane are more interesting and are depicted in Fig. \ref{MT}(c).  At 1.8 K, $M$ first increases linearly with increasing $B$ up to 2.2 T, where it shows a sharp jump representing a metamagnetic transition. It then increases with increasing $B$, but makes a cusp-like feature, followed by a linear increase before saturating above 6 T. The saturated moment at 1.8 K is  12.8 $\mu_B$ per formula unit. With increase in temperature, the metamagnetic transition remains in the entire
 temperature range below $T_N$. This transition, however, occurs at lower $B$ as the temperature increases. The magnetic saturation field also decreases with increasing temperatures. The cusp like feature, observed distinctly just above the metamagnetic transition at 1.8 K, gradually flattens above 100 K and is linear above 200 K. 
 
 Magnetization along the $c$-axis in YMn$_6$Sn$_4$Ge$_2$ has similar behavior as that in the parent compound, except that the  saturation is attained at much smaller field (8 T vs 13 T at 1.8 K). However, the magnetization of YMn$_6$Sn$_4$Ge$_2$ in the $ab$-plane has marked differences [see Figs. \ref{MT}(c and d)]. One similarity is that the metamagnetic transition at 1.8 K occurs at the same magnetic field of 2.2 T, but the behavior after the metamagnetic transition is quite different. While there is a cusp-like feature immediately after the metamagnetic transition in YMn$_6$Sn$_4$Ge$_2$, the parent compound has a linear $M$ dependence and is in the TCS magnetic phase. The cusp-like feature in the parent compound appears only in the FL phase between 6.7  T and 9.8 T (at 1.8 K). The FL phase in YMn$_6$Sn$_6$ is short lived and disappears above 170 K. 
 
 In Fig. \ref{MT}(d) we show the magnetic phase diagram of YMn$_6$Sn$_4$Ge$_2$ constructed by plotting the field derivative of magnetic moment ($dM/dB$) as a function of magnetic field in the temperature interval from 1.8 K to 350 K. The phase below the metamagnetic transition is the distorted spiral (DS) as this phase is similar to the DS phase of the parent compound [based on the magnetic structure determined in zero external magnetic field described in section \ref{sec:2}(D)]. The phase at magnetic saturation is the FF phase. We mark the intermediate  phases  as the unknown phases U1 and U2 because these phases are different from both the TCS and FL phase of the parent compound as discussed in more details in the following sections. 
 
 \subsection{Magnetotransport properties}
 
\begin{figure}[h]
\begin{center}
\includegraphics[width=1\linewidth]{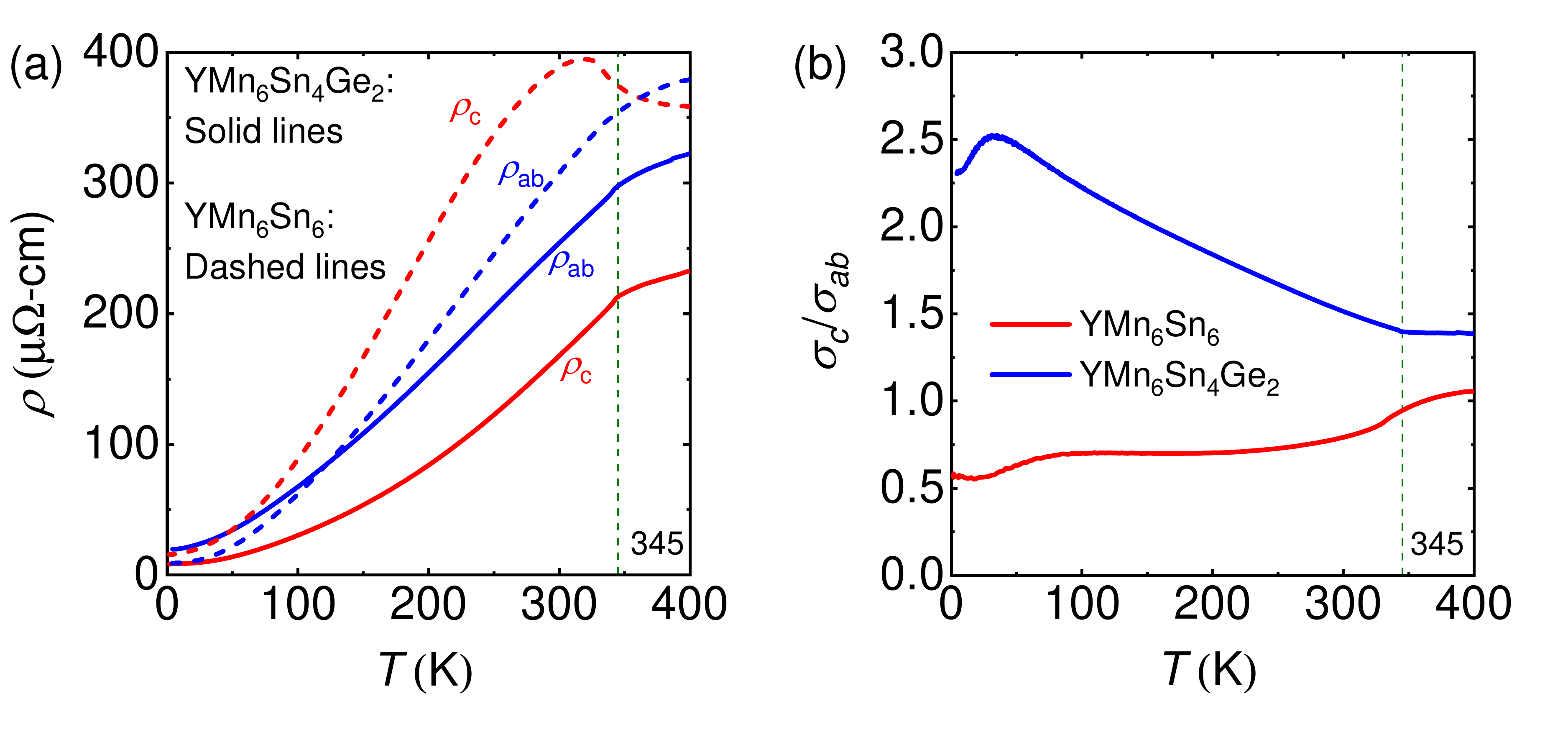}
\caption{\small (a) Electrical resistivity of YMn$_6$Sn$_4$Ge$_2$ (solid lines) and YMn$_6$Sn$_6$ (dashed lines)  as a function of temperature measured with the current applied along the $c$-axis ($\rho_c$) and in the $ab$-plane ($\rho_{ab}$). The YMn$_6$Sn$_6$ data are adopted from Ref. \cite{siegfried2022magnetization}. (b) Ratio of the $c$-axis to $ab$-plane conductivity of the two compounds.}
    \label{Rho}
    \end{center}
\end{figure}  

Electrical resistivitiy of YMn$_6$Sn$_4$Ge$_2$ measured with the electric current applied along the $c$-axis ($\rho_{c}$), and in the $ab$-plane ($\rho_{ab}$) is shown by solid red and blue lines, respectively, in Fig \ref{Rho}(a). In each direction, the resistivity decreases with decreasing temperature indicating the metallic behavior of the material in the entire temperature range from 400 to 1.8 K. Both $\rho_c$ and $\rho_{ab}$ show a marked kink at 345 K indicating the reduction of the spin scattering of the charge carriers with the onset of the magnetic ordering. It is consistent with the $T_N$ determined from susceptibility measurements in Fig. \ref{MT}(a). Resistivity for YMn$_6$Sn$_6$ is shown for comparison by the dashed lines in Fig. \ref{Rho}(a) and is very sensitive to the spiral ordering , especially when measured with the current along the $c$-axis \cite{siegfried2022magnetization}. The resistivity does not show any noticeable change when YMn$_6$Sn$_6$ orders into the commensurate antiferromagnetic phase at 345 K, but shows a remarkable change of slope at 333 K as it enters into the incommensurate spiral phase. The fact that a remarkable kink is observed in resistivity at 345 K in YMn$_6$Sn$_4$Ge$_2$ indicates that this compound directly orders into the incommensurate spiral state (which is verified by neutron diffraction, presented below). The $c$-axis resistivity, therefore, can provide important information about the nature of the magnetic ordering in this compound as it does in YMn$_6$Sn$_6$ \cite{siegfried2022magnetization}. Even more interesting is that the resistivity (in either direction) in YMn$_6$Sn$_4$Ge$_2$ is smaller than that in the parent compound, more significantly along the $c$-axis than in the $ab$-plane. It suggests that the $c$-axis conductivity is significantly enhanced by Ge doping of YMn$_6$Sn$_6$. This effect can be seen more clearly in  Fig. \ref{Rho}(b) where the ratio of $c$-axis to $ab$-plane conductivity ($\sigma_c/\sigma_{ab}$, where $\sigma$ = $1/\rho$) for both  YMn$_6$Sn$_6$ and YMn$_6$Sn$_4$Ge$_2$ are plotted between 1.8 K and 400 K. It can be seen that Ge doping, despite being isoelectronic doping, introduces a significant change in the fermiology. 

\begin{figure}[ht]
\begin{center}
\includegraphics[width=1\linewidth]{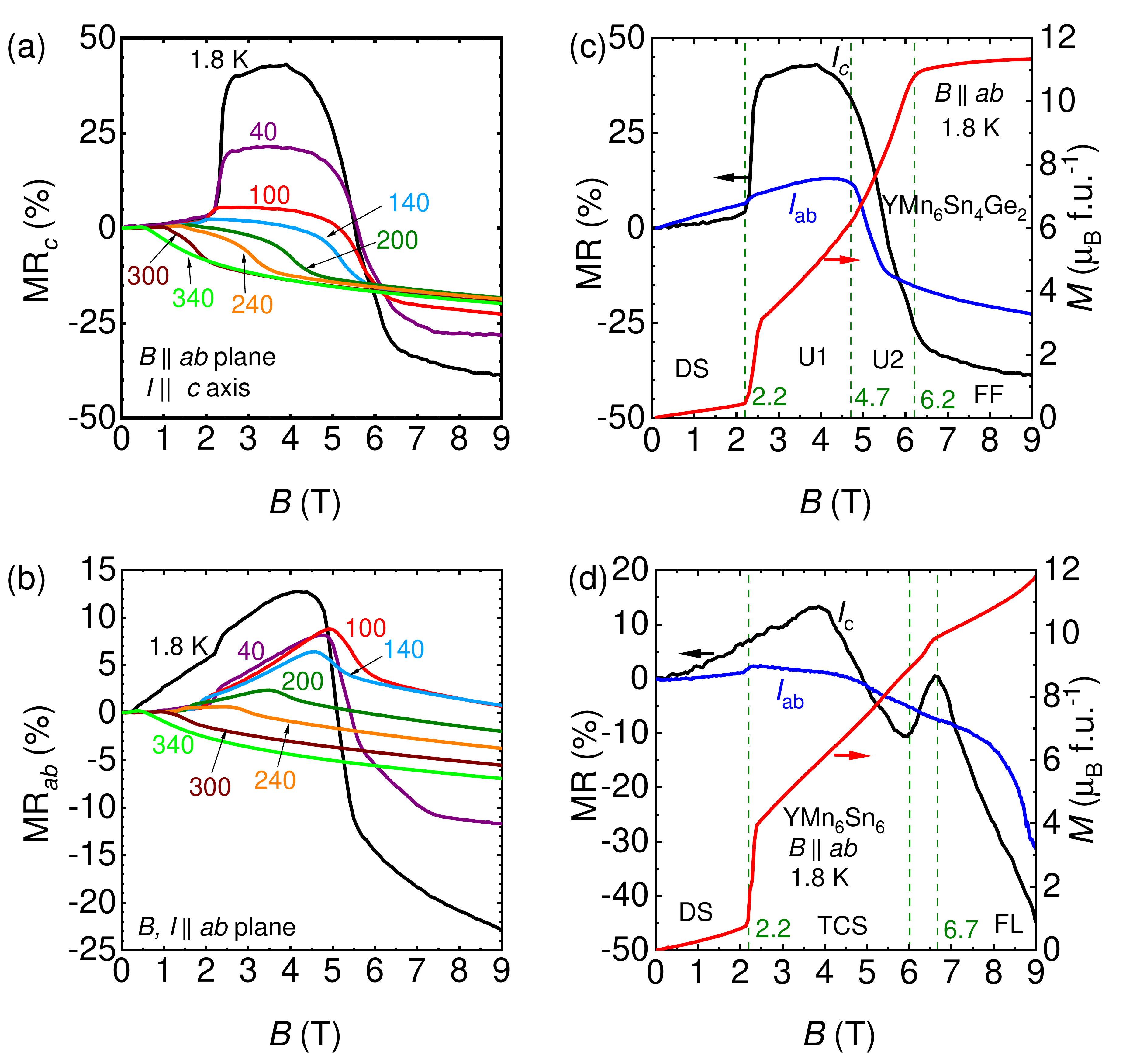}
\caption{\small Magnetoresistance of YMn$_6$Sn$_4$Ge$_2$ at selected temperatures measured with current applied along the (a) $c$-axis, and (b) in the $ab$-plane. Magnetoresistance (black and blue lines, left axis) and Magnetization (red line, right axis) of (c) YMn$_6$Sn$_4$Ge$_2$, and (d) YMn$_6$Sn$_6$ at 1.8 K. The YMn$_6$Sn$_6$ data in panel (d) are adopted from Ref. \cite{siegfried2022magnetization}. In each MR measurement, the magnetic field was applied in the $ab$-plane. In panels (c) and (d), the black (blue) curve represents the MR measured with current applied along the $c$-axis, $I_c$ ($ab$-plane, $I_{ab}$). The green dashed lines in panels (c) and (d) are a guide to eye for the indicated magnetic fields.}
    \label{MR}
    \end{center}
\end{figure} 

To understand the effect of the different magnetic phases on magnetotransport, we measured magnetoresistance (MR). The MR is defined by $(\rho_B-\rho_{0})/\rho_{0}\times 100 \%$  where $\rho_{B}$ $(\rho_{0})$ is resistivity in finite (zero) $B$. Let us first look at the $c$-axis MR (MR$_c$) which is depicted in Fig. \ref{MR}(a). At 1.8 K, MR$_c$ first increases slightly until it reaches the metamagnetic transition, where it jumps sharply by $\approx$ 42\%. With further increase in $B$ it shows a plateau until about 4.5 T where it decreases sharply and then flattens above 6.2 T where magnetic saturation is attained. On increasing temperature, the same MR behavior persists up to 200 K. At and above 240 K, the MR decreases with increasing magnetic field above the metamagnetic transition. The overall behavior of MR measured with current in the $ab$-plane (MR$_{ab}$) as depicted in Fig. \ref{MR}(b) is quite similar to that of MR$_{c}$, with the exception of the magnitude of jump at the metamagnetic transition, especially at lower temperatures. The MR$_{ab}$ jump at 1.8 K is very small compared to that of MR$_{c}$ as can be see in Fig. \ref{MR}(c). Here, the four magnetic phases inferred from the magnetization [red curve in Fig. \ref{MR}(c); also see Fig. \ref{MT}(d)] are clearly observed in the MR measurements in both directions. 

Comparison of MR and magnetization for an in-plane magnetic field at 1.8 K of YMn$_6$Sn$_4$Ge$_2$ and YMn$_6$Sn$_6$ can be seen in Figs. \ref{MR}(c) and (d), respectively. First, let us compare MR$_c$ [black curves in Figs. \ref{MR}(c) and (d)]. In YMn$_6$Sn$_6$, MR$_c$ changes very little, if at all, at the metamagnetic transition and shows a sharp drop at around 4 T due to the Lifshitz transition \cite{siegfried2022magnetization}. MR$_c$ then sharply increases before entering into the FL phase, then decreases continuously, and flattens after attaining magnetic saturation (not shown here). MR$_c$ of YMn$_6$Sn$_4$Ge$_2$, on the other hand, shows a significant jump at the metamagnetic transition, plateaus until about 4.5 T, above which it decreases rapidly, and then flattens after entering into the saturated state. This MR$_c$ behavior provides crucial information about the possible magnetic phases stabilized by the magnetic field in YMn$_6$Sn$_4$Ge$_2$. In our previous study of YMn$_6$Sn$_6$ 
\cite{siegfried2022magnetization}, we found that MR$_c$ is sensitive to the angle of rotation of spins between two consecutive planes. For example, this angle changes negligibly at the DS to TCS phase transition at the metamagnetic transition, but the change is larger at the TCS to FL phase transition. Thus, the  MR$_c$ shows negligible change during the former transition while it increases by about 10\% during the latter. Now, looking at MR$_c$ of YMn$_6$Sn$_4$Ge$_2$, the fact that it increases by about 42\% at the metamagnetic transition suggests that the inter-layer angle change between DS and U1 is much larger than that between the DS and TCS of the parent compound. The sharp drop in MR$_c$ at 4.7 T is consistent with a second magnetic phase transition to U2 inferred from the magnetization data that shows a sudden slope change at the same field. The forced ferromagnetic (FF) phase is attained above 6.2 T.     

\subsection{Neutron Diffraction: Magnetic structure in zero external magnetic field}
\begin{figure}[h]
\begin{center}
\includegraphics[width=0.98\linewidth]{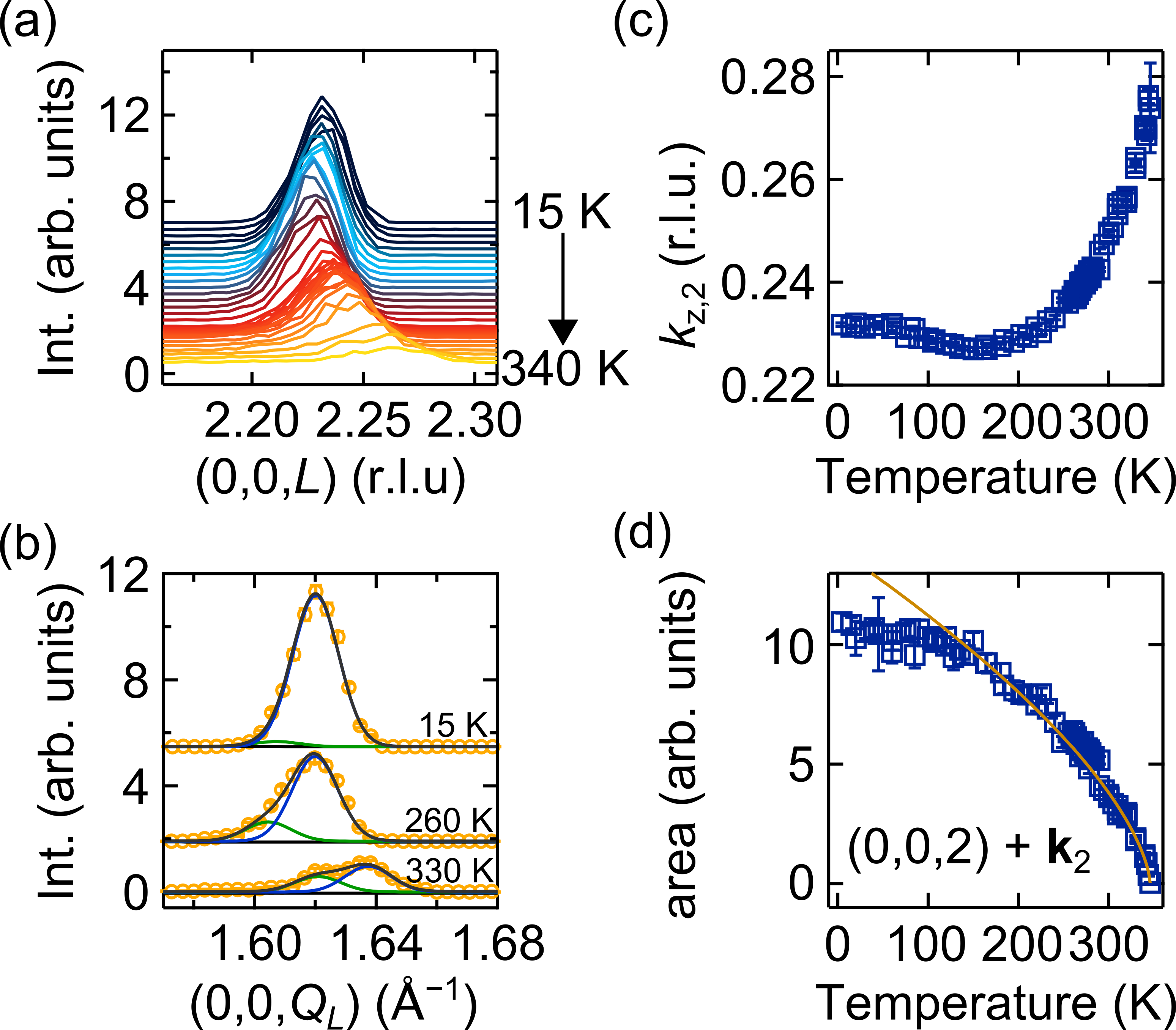}
    \caption{\small $L$-scans spanning the magnetic Bragg peaks at $(0,0,2) + \mathbf{k}_i$. (a) Scans between 15 K and 340 K, where scans are offset from one another along the intensity-axis by an amount proportional to the temperature. (b) Individual scans at 15 K, 260 K, and 330 K, offset along the Intensity-axis for clarity, with two-Gaussian fits to the data. The green solid line is the fit for the $\mathbf{k}_1$ magnetic peak, and the blue solid line is the fit for $\mathbf{k}_2$. The black and grey solid lines are for the linear background and the total fit, respectively. (c) The evolution of the $\mathbf{k}_2$ wavevector and (d) the area of the $(0,0,2) + \mathbf{k}_2$ magnetic peak with temperature. The solid orange line in (d) is a power law fit to the data.}
    \label{TAIPANdata}
    \end{center}
\end{figure}

The zero-field magnetic structure of YMn$_6$Sn$_6$ has a temperature dependent periodicity along the $c$-axis with wavevector $\mathbf{k} = (0, 0, k_z)$, where $k_z$ ranges from commensurate at $0.5$ reciprocal lattice units (r.l.u.) just below the onset of $T_N$ down to $0.26$ r.l.u.\ (incommensurate) at 4 K. YMn$_6$Sn$_4$Ge$_2$ is also incommensurate along the $c$-axis, however, there is no sign of the initial commensurate phase. The total change in pitch between the onset of magnetic order and the lowest temperature measured is much smaller than that in the parent compound ranging from $k_z=0.27$ r.l.u.\ at $T_N=345$ K to 0.23 r.l.u.\ at 4 K. Fig.~\ref{TAIPANdata} shows a summary of the results from the Taipan experiment, where the temperature evolution of the magnetic wavevector [Fig.~\ref{TAIPANdata}(c)] and area [Fig.~\ref{TAIPANdata}(d)] were tracked via $L$-scans across the $(0,0,2) + \mathbf{k}$ Bragg peak. At lower temperatures, a one-peak Gaussian fit is sufficient to analyze the data. However, at temperatures above $\approx 100$ K, the FWHM of the Bragg peak monotonically increases with increasing temperature, with values larger than that of the calculated instrumental resolution. Above $\approx 320$ K, it is clear that there are actually two magnetic Bragg peaks in close proximity to one another (similar to that in the parent compound). Fits comprised of two Gaussian peaks were therefore used to analyze the temperature dependent parameters. The FWHM of each Gaussian peak was fixed to be that of the calculated instrumental resolution for the $(0,0,2) + \mathbf{k}$ position (0.018 {\AA}$^{-1}$). The peak with the smaller-$q$ value is denoted by the wavevector, $\mathbf{k}_1$, and the peak with the larger-$q$ value is denoted by the wavevector, $\mathbf{k}_2$. The fits for representative data at 15, 260, and 330 K can be seen in Fig.~\ref{TAIPANdata}(b). The neutron data show that the magnetic structure associated with the $\mathbf{k}_2$ wavevector dominants than that belonging to $\mathbf{k}_1$, although it does appear that the $\mathbf{k}_1$ magnetic Bragg peak persists to the base temperature measured [fit results for the $\mathbf{k}_1$ structure are shown in Fig.~S1(c) and (d)]. This contrasts with the parent compound, where the volume of the $\mathbf{k}_1$ and $\mathbf{k}_2$ magnetic domains is approximately equal throughout the same temperature range.

The power law, $I = I_0 \left( 1-\frac{T}{T_N} \right) ^{2 \beta }$, was fit to the integrated area data for the Bragg peak, $(0,0,2)+\mathbf{k}_2$, shown as the solid line in Fig.~\ref{TAIPANdata}(d). The fit was performed using only data points near $T_N$, between 300 and 346 K, with $I_0 = 14.0 \pm 0.6$, $T_N = 345.95$ K $\pm 0.02$ K, and $\beta = 0.323 \pm 0.008$. The $\beta$ value extracted from the power law fit is $\approx \frac{1}{3}$, which is typical for various three-dimensional magnetic materials.

\begin{figure}[h]
\begin{center}
\includegraphics[width=0.95\linewidth]{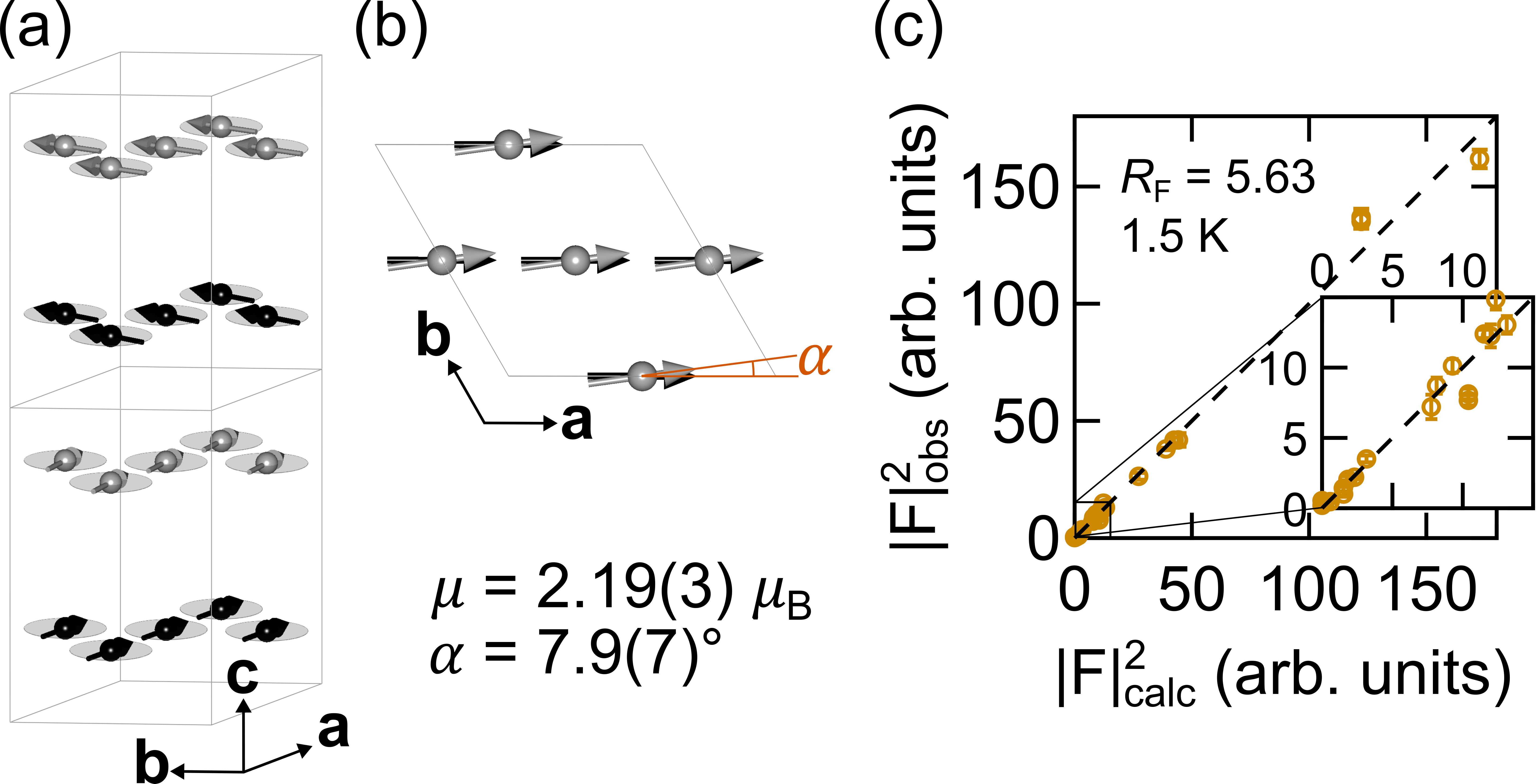}
    \caption{\small YMn$_6$Sn$_4$Ge$_2$ magnetic structure at 1.5 K. (a) View of the magnetic structure showing the moment rotation along the $c$-axis. Only Mn atoms are shown and the two layers of Mn atoms within a unit cell are shaded black and grey. The nuclear unit cell boundaries are outlined in grey. (b) View of the magnetic structure in the $ab$-plane of one unit cell. The Mn atoms within a unit cell are nearly ferromagnetically aligned, however, a small angle $\alpha=7.9(7)^{\circ}$ separates them. (c) The results of the magnetic structure refinement at 1.5 K, where the agreement between the observed ($\left| F \right| ^2 _{obs}$) and calculated ($\left| F \right| ^2 _{calc}$) structures factors is $R_F$-factor$ = 5.63$. }
    \label{magstr}
    \end{center}
\end{figure}

Data from the HB-3 triple-axis spectrometer were used to refine the nuclear [Fig.~S1(a)] and magnetic structure. Data were collected as $\theta - 2\theta$ scans across Bragg peaks, and the integrated area of Gaussian fits to the peaks were corrected for the Lorentz factor to extract the observed structure factor, $\left| F \right| ^2 _{obs}$. These values were used to refine the magnetic structure using the program, FULLPROF \cite{Rodriguez-carvajal1993}, and the results are shown in Fig.~\ref{magstr}. The refined structure is the staggered spiral where all moments within a Mn kagome plane are ferromagnetically ordered and constrained to the $ab$-plane. The moments between Mn planes (along the $c$-axis) are rotated relative to one another by an angle $\alpha$ within a unit cell (between Mn layers separated by Sn$_3$ block) and by an angle $\beta$ between unit cells (between the Mn layers separated by YSn$_2$ layer) [see Fig. \ref{structure}(a)]. The sum $\alpha + \beta$ determines the pitch length---and wavevector---of the spiral. At 1.5 K, $\mathbf{k}=(0,0,0.23)$ and the refined value $\alpha = 7.9(7)^{\circ}$, leading to a value of $\beta = 75.0(7)^{\circ}$. The total moment per Mn atom is 2.19(3) $\mu_B$. A smaller set of data were collected at 250 K, where the refinement yielded $\mu = $1.61(8) $\mu_B$ and $\alpha = 4(2)^{\circ}$, with an agreement $R_F$-factor$ = 10.6$ [refinement shown in Fig.~S1(b)]. The refined moment values at both 1.5 K and 250 K are in agreement with those from magnetization measurements [Figs.~\ref{MT}(b,c)].

\section{Discussion}
One marked difference between YMn$_6$Sn$_4$Ge$_2$ and YMn$_6$Sn$_6$ observed in neutron diffraction data at base temperature 4 K is in the spiral pitch, $k_z$, which is 0.23 in the former compound and 0.26 in the latter. As such, the refined values of the two turning angles, $\alpha$ and $\beta$, in YMn$_6$Sn$_4$Ge$_2$ are 7.9$^{\circ}$ and 75.0$^{\circ}$. These angles are quite different in the parent compound where $\alpha \approx -20^{\circ}$, and $\beta \approx  110^{\circ}$. This suggests that unlike the parent compound where $J_1<0$ and $J_2>0$, both of these exchange interactions are negative (i.e. $J_{1,2}<0$) in YMn$_6$Sn$_4$Ge$_2$. In fact, the ratio of the exchange interactions $J_2/J_1$ and $J_3/J_1$ can be calculated from $\alpha$ and $\beta$ using the following relations: \cite{ghimire2020competing,Rosenfeld2008a}
\begin{align*}
\alpha &  =-\mathrm{sign}(J_{1}J_{3})\cos^{-1}\left(  \frac{J_{2}J_{3}}%
{J_{1}^{2}}-\frac{J_{3}}{J_{2}}-\frac{J_{2}}{4J_{3}}\right),\\
\beta &  =\cos^{-1}\left(  \frac{J_{3}J_{1}}{J_{2}^{2}}-\frac{J_{1}}{4J_{3}%
}-\frac{J_{3}}{J_{1}}\right),\\
\alpha+\beta &  =\cos^{-1}\left(  \frac{J_{1}J_{2}}{8J_{3}^{2}}-\frac{J_{2}%
}{2J_{1}}-\frac{J_{1}}{2J_{2}}\right).
\end{align*}

Experimental values of $\alpha$, and $\beta$ yield $J_2/J_1$ = 0.142 and $J_3/J_1$ = -0.0692, confirming that $J_1$ and $J_2$ are both ferromagnetic in YMn$_6$Sn$_4$Ge$_2$. This means that replacing Sn by Ge in the 2c position reverses the sign of both $J_2$ and $J_3$. As one can anticipate, the effect of Ge substitution is the strongest regarding $J_2$, but the effect on $J_3$, while sign-changing, is small in terms of the absolute change. The change of sign of $J_2$ is a large and unexpected effect. Indeed, intercalating a magnetic ion such as Tb, in the same plane, leads to an effective ferromagnetic $J_1$ interaction, which is readily understood in terms of exchange coupling between Tb and Mn \cite{connor2022origin,Ke} (regardless of the sign of the Tb-Mn exchange coupling). Ge is nonmagnetic, so the sign change, on the first glance, seems rather mysterious.

\begin{figure}[h]
\begin{center}
\includegraphics[width=1\linewidth]{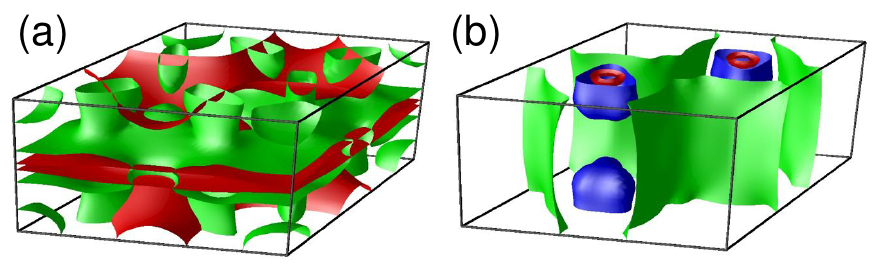}
\\
\includegraphics[width=1\linewidth]{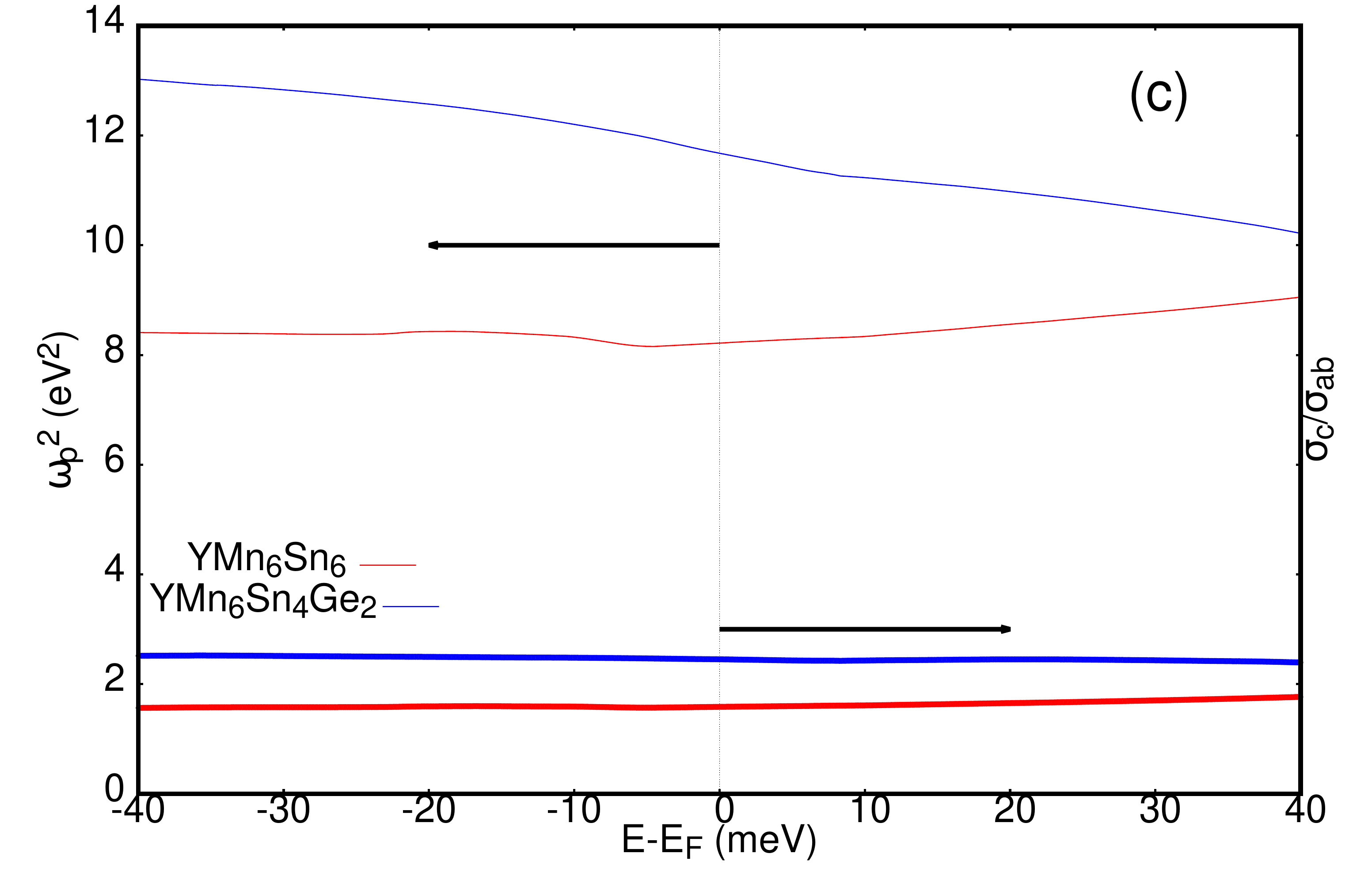}
    \caption{\small Calculated Fermi surface of ferromagnetic YMn$_6$Sn$_4$Ge$_2$ for (a) spin majority, and  (b) spin minority bands, (c) Calculated $c$-axis plasma frequency squared $(\omega_p^2)_z$ (left axis), and ratio of $c$-axis to $ab$-plane conductivity (right axis), for the Fermi surface depicted in panels (a) and (b).}
    \label{FS}
    \end{center}
\end{figure}

A useful hint is provided by our resistivity measurements, which show that Ge substitution increases the conductivity along the $z$ direction (where $z \parallel c$-axis here)  [Fig. \ref{Rho} (a)], thus making the material more metallic. This is consistent with the DFT calculations, which show the Fermi surface with quasi-2D sheets (in the spin-majority channel), which is absent in the parent compound [Figs.~\ref{FS}(a and b)] \cite{ghimire2020competing,siegfried2022magnetization}. Direct calculations of the plasma frequency squared $(\omega_p^2)$, which determines the conductivity in the single scattering rate approximation, also support this assertion; the calculated $(\omega_p^2)_z$ increases by 50\% upon Ge substitution as shown in Fig. \ref{FS}(c) (the in-plane conductivity also increases, albeit less, but is irrelevant for this discussion)! The calculated $\sigma_c/\sigma_{ab}$ presented in Fig. \ref{FS}(c) agrees well with the experimental data depicted in Fig. \ref{Rho}(b). Furthermore, better metallicity translates into a kinetic energy advantage for a ferromagnetic coupling, which is exactly what we deduced from our neutron measurements.

This provides insight into the role of  Ge substitution at the 2c position for both the magnetic ordering and fermiology of YMn$_6$Sn$_4$Ge$_2$ that change both the magnetic and transport properties measured in zero or small magnetic fields. These changes in magnetic and electronic structures are certainly responsible for the observed marked difference in the magnetic phases and the magnetoresistance behavior at higher magnetic field. Magnetic field dependent neutron diffraction experiments are required to determine the magnetic structure stabilized at the higher magnetic fields. More importantly, it will also be intriguing to perform several other measurements such as ARPES, STM \cite{li2022manipulation}, thermoelectric effect \cite{roychowdhury2022large}, and electromagnetic induction \cite{kitaori2021emergent} to understand the role of the altered magnetic and electronic structure in these various properties observed in the parent compound YMn$_6$Sn$_6$.

\section{Conclusion}\label{sec:4}

In summary, we have performed systematic transport, magnetic, and neutron diffraction experiments in a wide temperature and magnetic field range on single crystals of YMn$_6$Sn$_4$Ge$_2$. The experimental studies  have been supported by DFT calculations. Our magnetotransport measurements confirm that the magnetic structure of the parent compound YMn$_6$Sn$_6$ can be tuned through Ge doping. From neutron diffraction, we observed that this compounds orders into an incommensurate staggered spiral magnetic structure that persists from the N\'{e}el temperature to 1.5 K. An in-plane external magnetic field stabilizes two magnetic phases that are markedly different from those in the parent compound. Both experimental and calculated results reveal that Ge substitution onto the 2c site alters the signs of the interplanar exchange interactions $J_2$ and $J_3$ influencing the pitch of the staggered spiral. This change in magnetic structure influences the Fermi surface making the material more conductive along the $c$-axis.  Altogether, our detailed studies of the kagome magnet, YMn$_6$Sn$_4$Ge$_2$, provide important information that will be crucial in understanding the magnetotransport properties and the magnetic structures in the class of R166 kagome magnets that is currently attracting a great deal of interest for both magnetic and electronic topological states.  

\section{Methods}\label{S2}

Single crystals of YMn$_6$Sn$_4$Ge$_2$ were grown by the self-flux method using Sn as a flux. Y pieces (Alfa Aesar; 99.9\%), Mn pieces (Alfa Aesar; 99.95\%), Sn shots (Alfa Aesar; 99.999\%), and Ge pieces (Alfa Aesar; 99.9999\%)  were loaded into a 2-ml aluminum oxide crucible in a molar ratio of 1:6:18:2. The crucible was then sealed in a fused silica ampule under vacuum. The sealed ampule was heated to 1150$^{\circ}$C over 10 hours, kept at 1150$^{\circ}$C for 10 hours, and then cooled to 650$^{\circ}$C at the rate of 5$^{\circ}$C/h. Once the furnace reached 650$^{\circ}$C, the tube was centrifuged to separate the crystals in the crucible from the molten flux.  Several well-faceted hexagonal crystals [see the inset in Fig. \ref{Xray}(b) for an optical image of the crystals] up to 40 mg were obtained in the crucible.

The crystal structure was verified by Rietveld refinement \cite{Mccusker1999} of a powder x-ray diffraction pattern collected on a pulverized single crystal using a Rigaku Miniflex diffractometer. The Rietveld refinement was performed using the FULLPROF software \cite{Rodriguez-carvajal1993}.

DC magnetization, resistivity, and magnetoresistance measurements were performed in a Quantum Design Dynacool Physical Property Measurement System (PPMS) with a 9 T magnet. ACMS II  option was used in the same PPMS for DC magnetization measurements. Single crystals of YMn$_6$Sn$_4$Ge$_2$ were trimmed to adequate dimensions for electrical transport measurements. Crystals were oriented with the  [0,0,1] and [1,1,0] directions parallel to the applied field for the c-axis and ab-plane measurements. Resistivity and Hall measurements were performed using the 4-probe method. Pt wires of 25 $\mu$m were used for electrical contacts with contact resistances less than 30 Ohms. Contacts were affixed with Epotek H20E silver epoxy. An electric current of 2 mA was used for the electrical transport measurements. Contact misalignment in the magnetoresistance measurement was corrected by symmetrizing the measured data in positive and negative magnetic fields. The magnetic and magnetotransport data (with current along $c$-axis) presented here were measured on the same single crystal. First, magnetic properties were measured in both directions. Then the same crystal was polished to measure the magnetotransport measurements. The resistivity and magnetotransport data with current in the $ab$-plane were measured in a second crystal from the same growth batch after characterizing the magnetic properties. 
  
Neutron diffraction data were taken on the thermal triple-axis neutron spectrometer, TAIPAN, \cite{danilkin2007taipan} at the Australian Centre for Neutron Scattering. A PG(002) vertically focusing monochromator and a PG(002) flat analyzer were used at a wavelength of 2.345 {\AA}. S\"{o}ller slit collimators with full-width-at-half-maximum (FWHM) angular divergences of $30'$-$20'$-$20'$ were placed before the monochromator, before the sample, and after the sample. Data were taken between temperatures of 360 K and 4 K using a cryo-furnace with He exchange gas for temperatures below 290 K. The sample used was a 19.7 mg single crystal of YMn$_6$Sn$_4$Ge$_2$ oriented in the $(H,H,L)$ scattering plane. The sample mosaic was determined to be 0.37$^{\circ}$, reflective of a good quality crystal. Additional neutron diffraction data were taken on the thermal triple-axis neutron spectrometer, HB-3, at the High Flux Isotope Reactor at Oak Ridge National Laboratory. A PG(002) vertically focusing monochromator and a PG(002) flat analyzer were used at a wavelength of 2.359 {\AA}. S\"{o}ller slit collimators with FWHM angular divergences of $48'$-$20'$-$20'$-$70'$ were placed before the monochromator, before the sample, after the sample, and after the analyzer. The sample was the same single crystal used in the TAIPAN experiment, but the crystal was oriented in the $(H,0,L)$ scattering plane and placed in a cryostat (1.5 K - 300 K).

For the density functional (DFT) calculations we first used, a projector augmented wave basis as implemented in the Vienna ab initio simulation package (VASP) \cite{VASP1,VASP2}, for structure optimization, and then used an augmented plane wave code \textsc{WIEN2k} \cite{WIEN2, WIEN2} for Fermi surface analysis. In all cases a generalized gradient approximation for the exchange and correlation functional \cite{perdew1996generalized} was utilized. No LDA+U or other corrections beyond DFT were applied. Up to $11\times11\times 6$ k-point mesh (64 irreducible points) was used to structural optimization, and $48 \times 48\times 25$ for the Fermi surface and transport analyses.

Error bars displayed in plots and uncertainties listed throughout the manuscript represent plus and minus one standard deviation.

\begin{acknowledgments}
N.J.G and H.B. acknowledge the support from the NSF CAREER award DMR-2143903. Crystal growth part of the work at George Mason University was supported by the U.S. Department of Energy, Office of Science, Basic Energy Sciences, Materials Science and Engineering Division. I.I.M. acknowledges support from the U.S. Department of Energy through the grant No. DE-SC0021089. Research conducted at ORNL's High Flux Isotope Reactor was sponsored by the Scientific User Facilities Division, Office of Basic Energy Sciences, US Department of Energy. The authors would like to acknowledge the support from the Australian Centre for Neutron Scattering through proposal P9799. The identification of any commercial product or trade name does not imply endorsement or recommendation by the National Institute of Standards and Technology.
\end{acknowledgments}
\section*{References}
%

\widetext
\begin{center}
\pagebreak
\hspace{0pt}
\vfill
\textbf{\large Supplementary Material}
\vfill
\hspace{0pt}
\end{center}

\setcounter{equation}{0}
\setcounter{figure}{0}
\setcounter{table}{0}
\setcounter{page}{1}
\makeatletter
\renewcommand\thesection{SM\arabic{section}}
\renewcommand{\theequation}{S\arabic{equation}}
\renewcommand{\thetable}{S\arabic{table}}
\renewcommand\thefigure{S\arabic{figure}}
\renewcommand{\theHtable}{S\thetable}
\renewcommand{\theHfigure}{S\thefigure}

\section*{SM1.~~~~First principles calculations}

\begin{table}[ht!]
\caption{Calculated total energy for YMn$_6$Sn$_5$Ge.}\label{T1}
\begin{center}
\par%
\begin{tabular}
[c]{c@{\hspace{0.3cm}}c@{\hspace{0.3cm}}c@{\hspace{0.3cm}}c@{\hspace{0.3cm}}c@{\hspace{0.3cm}}c}\hline
 Position of Ge substitution  & Energy (eV)                     \\
 \hline
2e   &   -85.75591    \\
2d   &   -85.89259     \\
2c   &   -86.24354    \\

 \hline
\end{tabular}
\end{center}
\end{table}

\section*{SM2.~~~~Neutron Diffraction}

\begin{figure}[ht!]
\begin{center}
\includegraphics[width=0.6\linewidth]{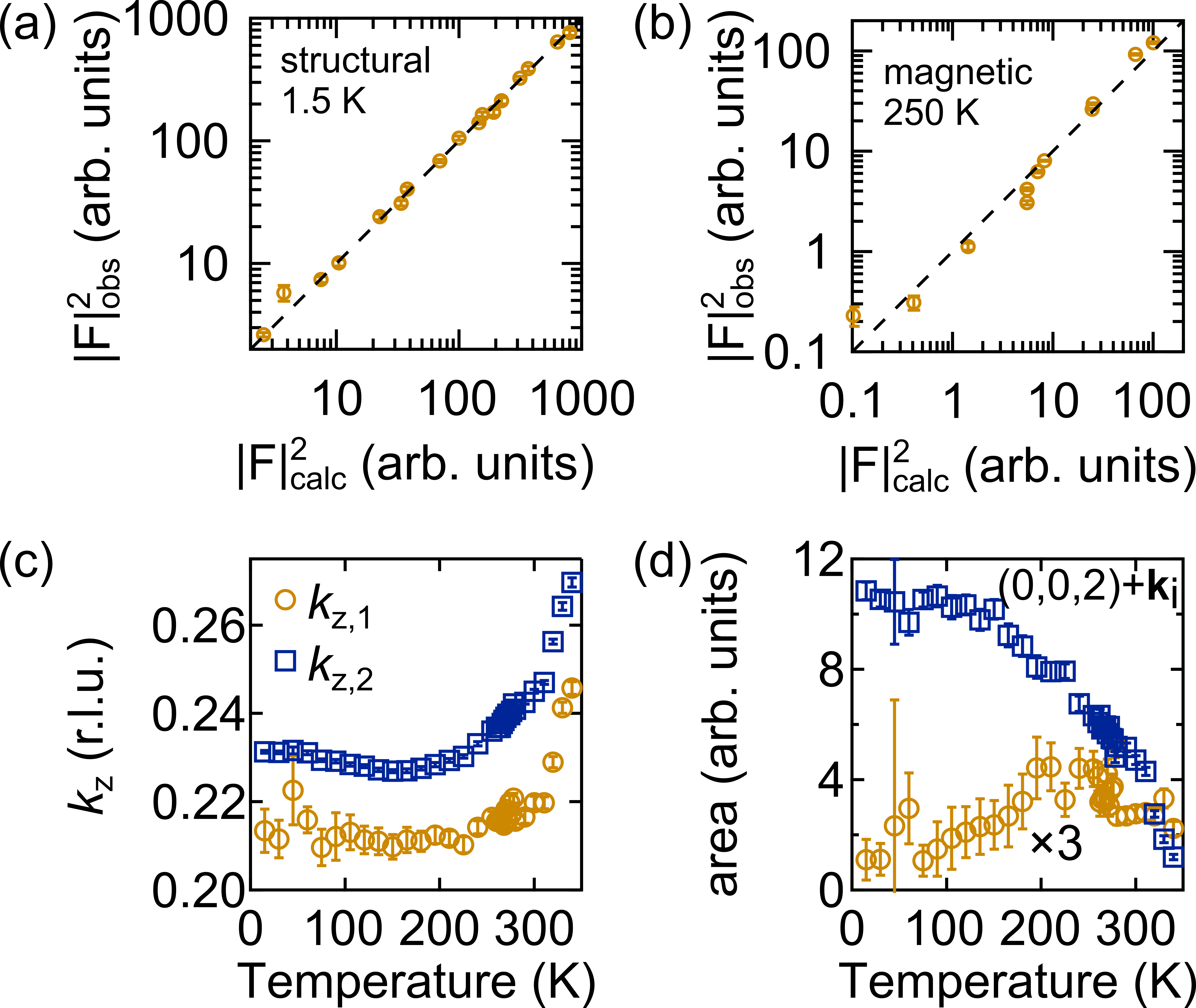}
    \caption{\small Rietveld refinement results from the HB-3 single crystal neutron diffraction experiment. (a) Data show the observed versus calculated structure factors for the YMn$_6$Sn$_4$Ge$_2$ nuclear structure at 1.5 K ($R_F$-factor $=2.65$). (b) Data show the observed versus calculated structure factors for the YMn$_6$Sn$_4$Ge$_2$ magnetic structure at 250 K ($R_F$-factor $=10.6$). The black dashed lines denote $\left| F \right| ^2 _{calc} = \left| F \right| ^2 _{obs}$. Temperature dependence of the (c) wavevector, $\mathbf{k}_i = (0,0,k_{z,i})$, and (d) area for the magnetic structures related to $i=1$ (orange open circles) and $i=2$ (blue open squares). The $(0,0,2) + \mathbf{k}_1$ area in (d) is multiplied by three for clarity.}
    \label{SIneutron1}
    \end{center}
\end{figure}

\end{document}